\documentclass[twocolumn,prl,aps,showpacs]{revtex4}
\usepackage{graphicx}
\usepackage{dcolumn}
\usepackage{bm}

\def\KNiF{K$_2$NiF$_4$}
\def\KCuF{K$_2$CuF$_4$}
\def\LaCuO{La$_2$CuO$_4$}

\def\CsAgF{Cs$_2$AgF$_4$}
\def\RbAgF{Rb$_2$AgF$_4$}

\begin{document}

\title{\textbf{2D Ferromagnetism in the High-T$_c$ Analogue 
Cs$_2$AgF$_4$}}
\author{
S.E.McLain,$^{1,2}$ 
D.A.Tennant,$^{3,4}$  
J.F.C.Turner,$^2$
T.Barnes,$^{5,6}$ 
M.R.Dolgos,$^2$
Th.Proffen$^7$,
B.C.Sales$^8$
and
R.I.Bewley$^1$
}
\affiliation{
$^1$ISIS Facility, Rutherford Appleton Laboratory, Chilton, Didcot OX11 0QX, UK\\
$^2$Department of Chemistry and Neutron Sciences Consortium, University of Tennessee, 
Knoxville, TN 37996, USA\\
$^3$Hahn-Meitner Institut, Glienicker Str. 18, Berlin D-14607, Germany 
$^4$School of Physics and Astronomy, University of St.Andrews, St.Andrews, Fife KY16 9SS, UK\\
$^5$Department of Physics and Astronomy and Neutron Sciences Consortium, 
University of Tennessee, Knoxville, Tennessee 37996, USA\\
$^6$Physics Division, Oak Ridge National Laboratory, Oak Ridge, TN 37831, USA\\
$^7$LANSCE, Los Alamos National Laboratory, Los Alamos, NM 87545, USA\\
$^8$Condensed Matter Sciences Division, Oak Ridge National Laboratory, Oak Ridge, TN 37831, USA 
}
\date{\today}
\pacs{75.10.Jm, 75.25.+z, 75.30.Ds, 75.40.Gb}

\begin{abstract}
Although the precise mechanism of high-T$_c$ superconductivity in the layered cuprates 
remains unknown, it is generally thought that strong 2D Heisenberg antiferromagnetism 
combined with disruptive hole doping is an essential aspect of the phenomenon.  
Intensive studies of other layered 3d transition metal systems have greatly extended 
our understanding of strongly correlated electron states, but to date have failed 
to show strong 2D antiferromagnetism or high-T$_c$ superconductivity.  For this reason 
the largely unexplored $4d^9$ Ag$^{\rm II}$ fluorides, which are structurally and perhaps 
magnetically similar to the $3d^9$ Cu$^{\rm II}$ cuprates, merit close study.  
Here we present a comprehensive study of magnetism in the layered Ag$^{\rm II}$ fluoride 
\CsAgF, using magnetic susceptometry, neutron diffraction and inelastic neutron 
scattering techniques.  We find that this material is well described as a 2D Heisenberg 
ferromagnet, in sharp contrast to the high-T$_c$ cuprates.  The exchange constant $J$ 
is the largest known for any material of this type.  We suggest that orbital ordering 
may be the origin of the ferromagnetism we observe in this material.
\end{abstract}

\maketitle

High-T$_c$ cuprates are distinctive in having [CuO$_2$] planes with very strong 
antiferromagnetic interactions between the spin-1/2 $3d^9$ Cu$^{\rm II}$ ions. 
Searches for antiferromagnetism with similarly large interactions in magnetic 
insulators containing other 3d spin-1/2 transition metal ions have not been 
successful to date.  
It may be instructive to extend the studies of magnetism in high-T$_c$ analogues in 
another direction, to materials possessing spin-1/2 4d electrons.  
The spin-1/2 $4d^9$ Ag$^{\rm II}$ ion is an obvious first choice for this 
research program, as it is the heavier congener of Cu$^{\rm II}$.  Cuprates also 
show charge-transfer character, and it is known that doped holes preferentially 
reside on the oxygen sites in the [CuO$_2$] planes rather than forming 
Cu$^{\rm III}$ ions.  In silver fluorides analogous charge-transfer is 
anticipated for fluorine sites in [AgF$_2$] planes \cite{Gro01}.

Despite having common formal oxidation states, Cu and Ag show marked differences in 
oxidation-state stability: although Cu$^{\rm I}$, Cu$^{\rm II}$ and Cu$^{\rm III}$
are all well represented in Cu chemistry, Ag$^{\rm I}$ and Ag$^{\rm III}$ dominate 
the solid-state and coordination chemistry of Ag \cite{Mul04,Sro97,Hou92}. 
To our knowledge, no oxide phases 
with [Ag$^{\rm II}$O$_2$] planes that are analogous to the cuprates exist, and 
AgO itself is best formulated as [Ag$^{\rm I}$Ag$^{\rm III}$O$_2$] \cite{Mul04}. 
Ag$^{\rm II}$ is an unusual oxidation state for Ag, although materials containing this ion 
in a fluoride lattice are known.

Ag$^{\rm II}$ is a powerful oxidizer, which presents practical difficulties and 
requires the use of reagents and solvents such as F$_2$ or HF.  However, using 
appropriate synthetic techniques, compounds containing Ag$^{\rm II}$F$_2$ square 
lattices can be prepared. In particular, ternary fluorides including \CsAgF\ and \RbAgF\ 
are known \cite{Ode74}. 
These compounds appear especially interesting as high-T$_c$ analogues, 
since they are structurally very similar to \KNiF\ and are related to the 
high-T$_c$ precursor 
\LaCuO; both families contain planes of [AgF$_2$] or [CuO$_2$] separated by planes of 
Cs/RbF or LaO.  Few studies of these Ag$^{\rm II}$ compounds have been reported to date, 
and their magnetic properties are not well understood \cite{Gro01,Ode74}. 
Here we report detailed 
measurements of the magnetic properties of \CsAgF, and compare with results for 
related materials.

The inverse of the measured molar static magnetic susceptibility of polycrystalline 
\CsAgF\ (Methods) is shown in Figure 1.  Our results are consistent with and extend 
the earlier data of Odenthal {\it et al.} \cite{Ode74}.
Note that the qualitative behaviour is 
characteristic of a ferromagnet rather than an antiferromagnet such as \LaCuO.  

\begin{figure}[ht]
\begin{center}
\resizebox{7cm}{!}{\includegraphics{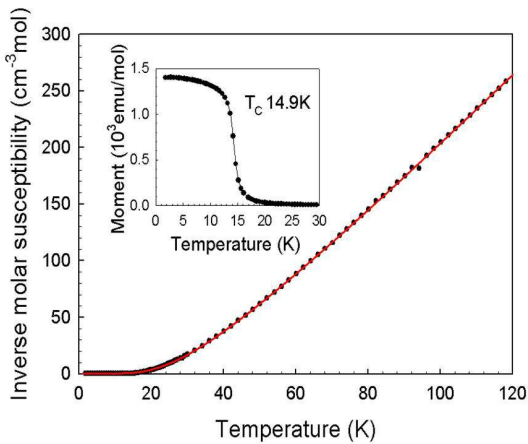}}
\caption{ 
Figure 1: The inverse of our measured molar susceptibility of \CsAgF\ (points) 
after background subtraction. A fit to the theoretical susceptibility of the 
2D spin-1/2 Heisenberg ferromagnet 
(the 10$^{th}$ order series of Baker {\it et al.} \cite{Bak67}) 
is shown as a solid line, and gives $g = 1.832$ and $J = -3.793$~meV.  
The insert shows the 15~K Curie transition and the low-temperature magnetic 
moment (with a simple interpolating curve).}
\label{fig1}
\end{center}
\end{figure}

The structural and electronic similarity to \LaCuO\ had suggested modelling magnetism in 
\CsAgF\ with the 2D square lattice spin-1/2 Heisenberg antiferromagnet (HSqL), 
described by the Hamiltonian 
\begin{equation}
H =   J\; \sum_{\langle {ij}\rangle }  
{\vec S}_i \cdot {\vec S}_j
\label{Hamiltonian}
\end{equation}
where $J > 0$ is the exchange constant.  We find that this model actually does 
describe the susceptibility of \CsAgF\ rather well over a wide range of 
temperatures, {\it albeit} with a ferromagnetic exchange constant $(J < 0)$.  
Our fit to the susceptibility is shown in Figure 1; this gave the parameters 
$g = 1.832$ and $J = -3.793$~meV.  We also considered 1D and 3D Heisenberg models, 
but these were found to be in clear disagreement with the measured susceptibility. 

The Ag$^{\rm II}$ ion moment was estimated from a field-magnetization scan to be 
0.8~$\mu_B$ at 5~K, corresponding to $g = 1.6$; a value of 1.0~$\mu_B$ would be expected 
for an isolated, isotropic spin-1/2 ion.  Note that in a magnetization measurement 
the extracted ionic moment would be lowered by the presence of non-magnetic 
impurity phases.  Covalency effects may also have lowered the moment in the pure material.  

The predominant difference between Ag$^{\rm II}$ and Cu$^{\rm II}$ is the increase in 
the principle quantum number of the valence orbitals. In this ternary silver fluoride, 
the combination of lower Coulomb repulsion of $4d^9$ compared to $3d^9$ electrons 
and charge transfer properties anticipated for Ag$^{\rm II}$F$_2$, which should 
differ from those of Cu$^{\rm II}$F$_2$, may lead to new effects due to orbital 
fluctuations.  The reduced $g$-value may be evidence for fluctuation or 
hybridization phenomena. 

A ferromagnetic ordering (Curie) transition was observed in \CsAgF\ near 
15~K (see Figure 1 insert).  Since an ideal 2D Heisenberg magnet 
only undergoes ordering at absolute zero \cite{Mer66}, this 15~K transition signifies the 
presence of additional interactions such as anisotropies or interlayer coupling.

Although the bulk susceptibility is useful for characterizing the general magnetic 
behaviour of a material, establishing the nature of these interactions at the 
atomic scale requires a local probe.  Inelastic neutron scattering is ideal for 
this purpose, as the magnetic interaction strengths and pathways can be inferred 
from the energy and momentum transfer to the sample.  As the impurities in the sample 
have very different magnetic properties than \CsAgF, it is reasonable to assume 
that they will not interfere with the interpretation of the inelastic neutron 
scattering signal.

Figure 2 shows the results of our inelastic neutron scattering measurements.  
Figure 2(a) shows the scattering from \CsAgF\ over the wavevector range 
$Q = 0.2 - 1.2$\AA$^{-1}$ at T = 8~K after background subtraction.  
Production of magnetic excitations is clearly observed above 6~meV.  The magnetic 
nature of these excitations was confirmed by a decrease in intensity 
and broadening of the signal upon heating to 35~K.  For comparison, Figure 2(b) 
shows the scattering predicted by the ferromagnetic 2D spin-1/2 Heisenberg model 
with $J = -5.0$~meV. 

\begin{figure}[ht]
\begin{center}
\resizebox{6cm}{9cm}{\includegraphics{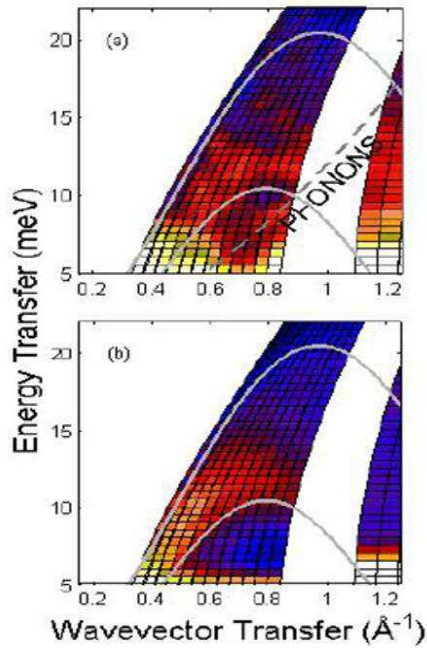}}
\caption{ 
Figure 2: (a) Low-wavevector inelastic neutron scattering from \CsAgF\ at 8~K, using 
E$_i = 100$~meV, corrected as described in text.  The intensity of scattering is 
indicated by the colour scale.  Below an energy transfer of 6~meV, incoherent 
scattering processes are dominant.  Magnetic excitations are observed from 
$0.4-1$ \AA$^{-1}$ and $6-15$~meV.  Solid grey lines bound the region in which 
appreciable scattering from magnetic excitations is expected.  The grey dashed 
curve indicates the onset of significant phonon scattering relative to the weak 
ferromagnetic scattering (above $\sim 1$ \AA$^{-1}$).  
(b) Simulated scattering for a 2D spin-1/2 Heisenberg ferromagnet, using an 
exchange constant of $J = -5.0$~meV.
}
\label{fig2}
\end{center}
\end{figure}

To determine $J$, scans in energy at constant wavevector were extracted and 
the peak positions were determined by fitting Gaussian profiles to these data.  
The data and fits are shown in Figure 3 (main panel).  The dispersion relation 
for ferromagnons in the 2D Heisenberg model was fitted to this data after 
correcting for instrumental effects, yielding $J = -5.0(4)$~meV for the exchange 
constant in \CsAgF.  (Recall that the fit to the susceptibility gave a somewhat 
lower value of $J = -3.8$~meV.)  For wavevectors above $\sim 1 $\AA$^{-1}$, scattering 
from phonons becomes dominant, so the higher-$Q$ data was not fitted.  
Below 6~meV, quasi-elastic incoherent scattering masks any magnetic signal 
that may be present.

\begin{figure}[ht]
\begin{center}
\resizebox{7cm}{!}{\includegraphics{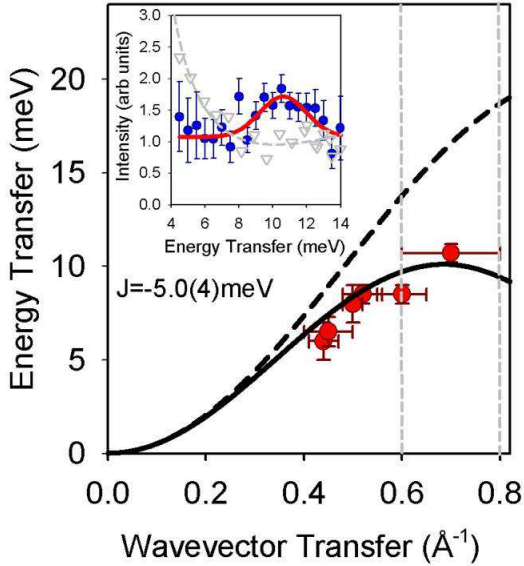}}
\caption{ 
Figure 3: (inset) Inelastic neutron scattering from \CsAgF\ with E$_i = 50$~meV, 
at temperatures of 8~K (blue circles) and 35~K (grey triangles), showing the range 
$0.6-0.8$ \AA$^{-1}$.  The line through the data is a fit as described in the text. 
(main panel) Fitted positions of scattering in wavevector and energy are plotted, 
corrected for resolution.  Dynamical scattering for the ferromagnetic powder lies 
in a continuum of values bounded by the lower solid and upper dashed curves. 
}
\label{fig3}
\end{center}
\end{figure}


The ferromagnetism we observe in \CsAgF\ is in striking contrast to the 
antiferromagnetism in \LaCuO, and is instead reminiscent of \KCuF.  This material 
is one of the few known 2D spin-1/2 ferromagnetic insulators \cite{Mou76,Li87}, 
and like \CsAgF\ is a 
Heisenberg ferromagnet, with $J = -1.0$~meV and T$_{\rm Curie} = 6.25$~K \cite{Li87}. 
The ferromagnetism in \KCuF\ has been attributed to a cooperative Jahn-Teller 
distortion and associated orbital ordering \cite{Hid83,Kho73}, as depicted in Figure 4(c).

According to the Jahn-Teller theorem \cite{Jah37}, 
the stereochemistry of a $d^9$ metal ion should 
distort so as to lower both the symmetry and the energy of the system.  
For a hypothetical Ag$^{\rm II}$F$_6$ octahedron, in which the Ag$^{\rm II}$ ion 
is in a $t_{2g}^6e{_g}{^3}$ electronic state, two structural pathways are 
commonly observed for this distortion \cite{Alb85}. Axial compression of the octahedron 
results in a partially filled $a_{1g}$ $(\sim 3z^2-r^2)$ ground state orbital, 
whereas axial elongation results in a $b_{1g}$ $(\sim x^2-y^2)$ ground state \cite{Alb85,Koo00}. 
Both distortions give rise to the same $D_{4h}$ local point group, which is 
the local symmetry of the Ag$^{\rm II}$ ion in the space group $I4/mmm$.  

There is, however, no known example of a system in which the electronic ground state is 
$a_{1g}$ $(\sim 3z^2-r^2)$ \cite{Kho05}; rather, systems that show axial compression are better 
described using a partially filled $(z^2-x^2)$ and $(z^2-y^2)$ basis 
on the metal center \cite{Kho73,Koo00}. 
The magnitude and precise nature of the distortion will depend, 
{\it inter alia}, on the axial electrostatic potential and anisotropic elastic 
constants for axial {\it versus} equatorial ligand displacement \cite{Gar04}, but the orbital 
occupancies and symmetries themselves will conform to this general model.  
In the square lattice copper halides \KCuF\ \cite{Li87} and (C$_2$H$_5$NH$_3$)$_2$CuCl$_4$
\cite{Mor94}, $a_{1g}$ and $b_{1g}$ orbital states couple to each other and to the electron 
spin and lattice degrees of freedom, producing a complex nonlinear Hamiltonian 
that involves all these degrees of freedom \cite{Mos04}. An antiferrodistortive pattern of 
$(\sim z^2-x^2)$ and $(\sim z^2-y^2)$ hole orbitals results, giving a strong ferromagnetic 
state.  In this state, the fluorine atoms are displaced from their 
symmetric $I4/mmm$ positions, 
lowering the symmetry to $Bbcm$.  We suggest that a similar mechanism 
is responsible for the observed ferromagnetism in \CsAgF.  If the difference in the 
exchange constant $J$ measured above and below T$_c$ is significant ($-5$~meV 
from neutron scattering versus $-3.8$~meV from the susceptibility), it may be 
associated with this orbital ordering transition.

Orbital ordering has been observed in other systems; cooperative orbital order in 
pseudo-cubic KCuF$_3$ leads to strong 1D antiferromagnetism, and strong 
antiferromagnetism has also been reported in KAgF$_3$, in which evidence for 
conduction above 50~K has been reported \cite{Gro03}.
However the absence of strong 2D antiferromagnetism among the copper halides in general
suggests that the arrangement of pseudo-octahedra in the perovskites or
associated Ruddlesden-Popper phases is important.

In a search for evidence of this structural distortion in \CsAgF, we carried out 
neutron diffraction measurements on \CsAgF\ at 6~K and 298~K.  The data were fit 
using the Rietveld method assuming the space groups $I4/mmm$ and $Bbcm$, the 
latter being the space group of the orbitally ordered phase of \KCuF\ \cite{Hid83}. 
As no significant difference in the quality of fit was observed between the $I4/mmm$- 
and $Bbcm$-symmetry models, our results are consistent with either space group.  
However, we note that the atomic displacement parameters for the equatorial 
fluorine atoms in the AgF$_6$ pseudo-octahedra are large, which may be a signature 
of disorder in the plane; the resolution of our diffraction experiments does not 
allow discrimination of this disorder.  This may be due to the limitation of 
statistics for a highly absorbing sample, in which small shifts in position 
may not be easily detectable.  These diffraction measurements showed that the 
AgF$_6$ pseudo-octahedra in \CsAgF\ are compressed, which is inconsistent 
with the $b_{1g}$ ground state familiar in \LaCuO; in that case the CuO$_6$ 
pseudo-octahedra show axial extension (see Figure 4(a)).  Future investigations 
are planned, using high resolution neutron diffraction and resonant X-ray diffraction
\cite{Man03,Bin04}, 
which should clarify the precise nature of the orbital ordering.

\begin{figure}[ht]
\begin{center}
\resizebox{7cm}{!}{\includegraphics{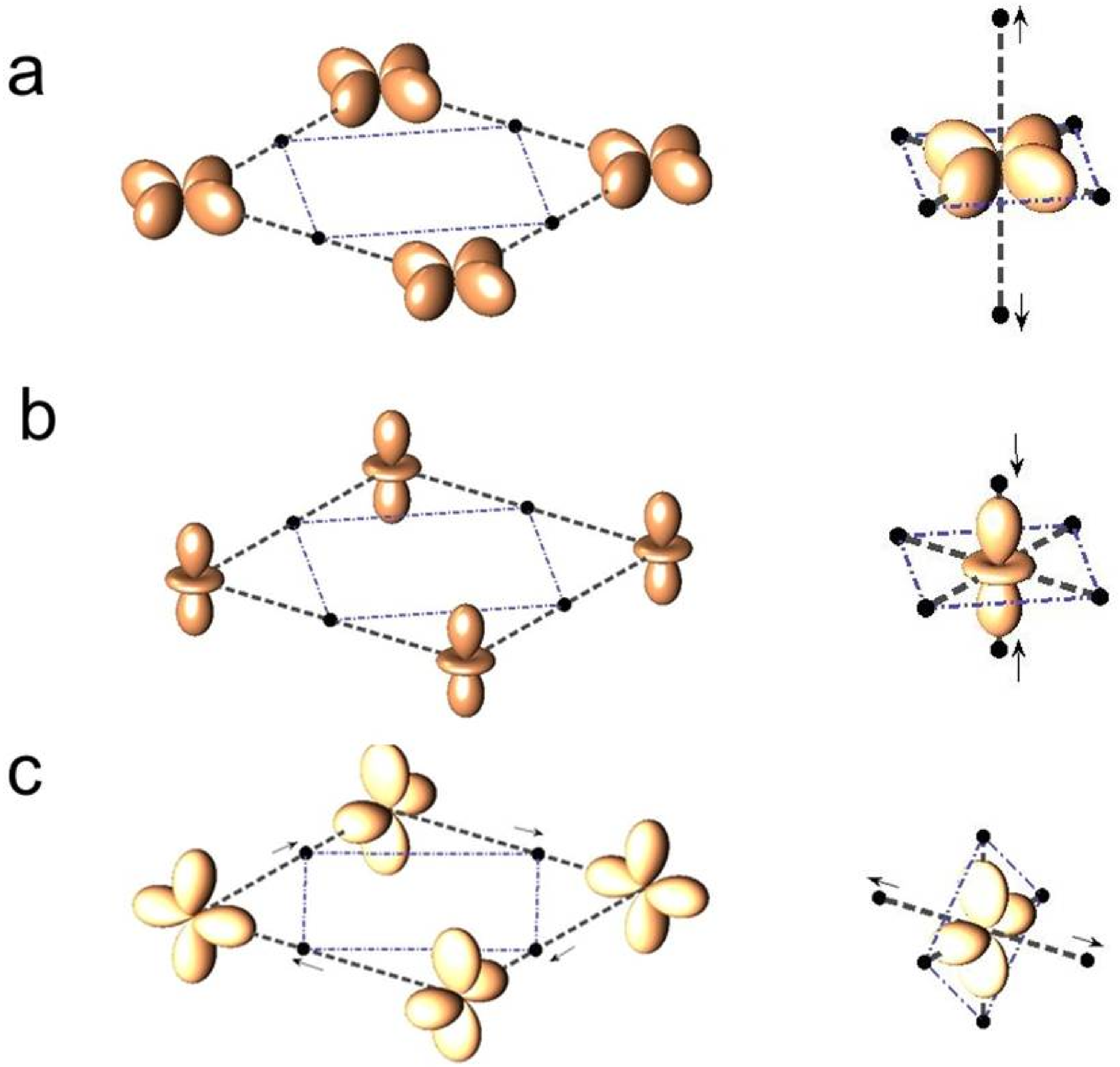}}
\caption{ 
Figure 4: Orbital ordering scenarios for Jahn-Teller active $3d^9$ and $4d^9$ 
layered perovskites. (a) Formation of a $b_{1g}$ ground state with partially 
filled $\sim x^2-y^2$ orbitals is accompanied by axial extension of octahedral 
ligands (right).  Large in-plane overlap with ligand $p$-orbitals gives very 
strong antiferromagnetism, as is found in the high-T$_c$ cuprates.  
(b) Formation of $a_{1g}$ ground states (partial filled $\sim 3z^2-r^2$ orbitals) 
is accompanied by axial compression (right). Weak 2D ferromagnetism in the layers 
or 1D behaviour between layers may occur.  (c) Anharmonic coupling to ligands 
favours the formation of staggered orbital ordering of $\sim z^2-x^2$ and 
$\sim z^2-y^2$ orbitals, as in \KCuF.  In-plane ligands are displaced from 
central positions, lowering the crystal symmetry. Strong ferromagnetic 
superexchange occurs through the unfilled in-plane orbitals. 
}
\label{fig4}
\end{center}
\end{figure}

The different $d$-orbital orientations proposed for \CsAgF\ and \LaCuO\ may have a simple 
origin in the electrostatics of these materials.  In the cuprates, the [CuO$_2$] 
layers possess a net negative charge, having two electrons per formula unit.  
For this reason it is energetically favourable for the positively charged holes to 
lie in the $ab$ plane, localized in $b_{1g}$ orbitals.  In contrast the [AgF$_2$] 
layers in \CsAgF\ are charge neutral, so there is no strong electrostatic 
orientation preference.  
($a_{1g}$ states are expected to be slightly preferred \cite{Gar04}.) 
To test the stability of the assumed $a_{1g}$ configuration of \CsAgF, we have 
also synthesized the isostructural \RbAgF\ \cite{Ode74}. 
This material shows magnetic properties 
very similar to \CsAgF, demonstrating that this magneto-orbital configuration 
is robust.  We speculate that charge-doping of Ag$^{\rm II}$ fluorides may alter 
the electrostatic forces sufficiently to allow in-plane holes and stabilize 
a strong antiferromagnetic state. Our initial investigation of electron-doped 
Cs$_{2-x}$Ba$_x$AgF$_4$ for $0 < x < 0.3$ however has not identified 
an antiferromagnetic state \cite{Dol04}.

The structural similarity to the high-T$_c$ precursor \LaCuO, combined with our 
confirmation of strong magnetic interactions between the Ag$^{\rm II}$ ions in 
\CsAgF, suggests the possibility of a rich variety of magnetic phases in 
Ag$^{\rm II}$ fluorides \cite{Gro01}. Given the oxidizing power of Ag$^{\rm II}$, it is 
unlikely that other non-fluoride anion lattices will accommodate this species; 
accordingly, we are currently exploring ternary and quaternary 
Ag$^{\rm II}$-fluoride Ruddlesden-Popper phases \cite{Dol04}. High pressure experiments, 
which have proved fruitful in other systems \cite{Mor94}, 
may also be particularly interesting.

In summary, we have found that the layered silver fluoride \CsAgF\ is well 
described magnetically as a 2D spin-1/2 Heisenberg ferromagnet, and the 
exchange constant $J$ is the largest known for any magnetic material of 
this type.  These conclusions followed from measurements of both static 
(susceptibility) and dynamic (inelastic neutron scattering) magnetic 
properties of polycrystalline \CsAgF. 

\section{Acknowledgements}

We would like to thank J. A. del Toro and the staff at the University of Liverpool 
for the use of their SQUID magnetometer, D. Scalapino, D. I. Khomskii and D. J. Singh 
for critical readings of the manuscript, and D. Argyriou and S.M. Bennington for 
useful discussions.  S. E. McLain acknowledges support from the U.S. National 
Science Foundation through award OISE-0404938.  Financial support was also provided 
by the EU through the Human Potential Programme under IHP-ARI contract 
HPRI-CT-1999-00020, the Manuel Lujan Neutron Scattering Center, funded by the 
U.S. Department of Energy Office of Basic Energy Sciences, and Los Alamos National 
Laboratory, funded by the U.S. Department of Energy under contract W-7405-ENG-36.  
J. F. C. Turner acknowledges the financial support of the U.S. National Science 
Foundation, through a CAREER award (Grant No. CHE 0349010), and the University of 
Tennessee though the Neutron Sciences Consortium.

\section{Methods}

\CsAgF\ was prepared through a solid state reaction between AgF$_2$ and CsF 
(Fluorochem USA).  In a typical preparation, 0.0178 mol of AgF$_2$ and 0.0356 mol 
of CsF were ground together in a dry box in an inert atmosphere of argon or 
N$_2$ using an agate mortar and pestle until they formed a visually homogenous 
mixture.  The mixture was then transferred to a pure gold reaction tube 
(height 9~cm, i.d. 1.5~cm) and placed inside a Schlenk flask and 
subsequently heated to 270$^o$C under a flow of argon for 24 hours.  The resulting 
lilac coloured solid was stored in an N$_2$-filled glove box in a desiccator over 
NaK$_3$ alloy.  Sample purity was checked by X-ray powder diffraction using a 
Bruker AXS Smart 1000 diffractometer equipped with a graphite monochromatized Mo-Ka 
source. The observed pattern and positions of diffraction rings corresponded 
to the \CsAgF\ unit cell parameters reported previously \cite{Ode74}. 
X-ray diffraction experiments at NSLS also confirmed the nature of the phases.

For the SQUID measurements, approximately 100~mg of \CsAgF\ powder was loaded 
into prefluorinated PTFE sample holders under an argon atmosphere.  
Measurements were made using a Quantum Design SQUID magnetometer, 
under a weak applied field of 100~Oe.  For the neutron diffraction measurements, 
approximately 0.5~g of \CsAgF\ powder was loaded into flame-dried quartz tubing 
(o.d. 0.4~cm, i.d. 0.30~cm), also under an argon atmosphere and subsequently 
flame-sealed under vacuum ({\it ca.} $10^{-3}$~mbar).  Measurements were performed 
using the Neutron Powder Diffractometer (NPDF) at the Manuel Lujan Neutron Scattering 
Center at Los Alamos National Laboratory.  All diffraction data were corrected for 
background scattering, incident neutron flux, multiple scattering and absorption 
effects, using the PDFgetN analysis procedures \cite{Pet00}. The data were analysed using 
reciprocal space Rietveld refinements, performed with the GSAS package \cite{Lar86}, and 
yielded the tetragonal lattice parameters 
$a = b = 4.5573(14)$ \AA \ and $c = 14.166(6)$ \AA \ at 6~K, and 
$a = b = 4.5905(18)$ \AA \ and $c = 14.213(6)$ \AA \ at 295~K, consistent with 
previous reports \cite{Ode74}. Multiphase Rietveld refinement of these data showed the presence 
of small quantities of AgF and CsF at the level of a few mole percent. 

For the inelastic neutron scattering measurements, approximately 10~g of 
\CsAgF\ powder were loaded into scattering cells under an argon atmosphere.  
As the neutron absorption cross-section for Ag is quite large (63.3~barns), 
a thin plate Al sample cell (3~mm width) was utilized.  The sample was initially 
placed in an envelope of prefluorinated PTFE sheet (5~cm x 5~cm x 0.01~mm) 
to prevent a chemical reaction with the aluminium sample cell.  Measurements were 
performed using the High Energy Transfer (HET) time-of-flight spectrometer at the 
ISIS pulsed neutron source, Rutherford Appleton Laboratory, UK.  HET is a direct 
geometry chopper spectrometer with large angular $^3$He detector coverage, from 
scattering angles of $\phi = 3^o$ to $7^o$ (4~m) and $\phi = 9^o$ to $29^o$ (2.5~m) 
in two separate banks, which makes it especially effective for the study of neutron 
scattering from powders.  Incident energies of 50 and 100~meV were selected and 
measurements made at temperatures of 8~K, 35~K and 295~K. Typical measurement times 
were 10~hours per run at an average proton current of $1700~\mu A$. Empty cell runs 
were also performed at the same energies and temperatures. All data were 
corrected individually for incident neutron flux and detector efficiencies.  
Empty cell runs were then corrected for sample absorption effects before 
being subtracted from the data sets obtained from the powder sample itself.  
This subtraction was necessary because scattering from PTFE remains significant 
at the low wave-vectors associated with magnetic excitations.

\end{document}